%% file: TPS_MAIN.tex
\documentclass[conference]{IEEEtran}
\IEEEoverridecommandlockouts
\usepackage{cite}
\usepackage{amsmath,amssymb,amsfonts}
\usepackage{algorithmic}
\usepackage{graphicx}
\usepackage{textcomp}
\usepackage{xcolor}
\usepackage{url}
\usepackage[draft=false]{hyperref}
\usepackage{comment}

\def\BibTeX{{\rm B\kern-.05em{\sc i\kern-.025em b}\kern-.08em
    T\kern-.1667em\lower.7ex\hbox{E}\kern-.125emX}}
\usepackage{multirow}
\usepackage{graphicx}
\usepackage{tabularx}

\begin{document}

\title{The Evolving Path of ``the Right to Be Left Alone'' - When Privacy Meets Technology\\
{\footnotesize \thanks{The views and opinions expressed in this paper are those of the author and do not necessarily reflect the official policy or position of Banca d'Italia.}}
}

\author{\IEEEauthorblockN{Michela Iezzi}
\IEEEauthorblockA{\textit{ICT Department} \\
\textit{Bank of Italy}\\
Rome, Italy \\
michela.iezzi@bancaditalia.it}}


\maketitle

\begin{abstract}
This paper deals with the hot, evergreen topic of the relationship between privacy and technology. We give extensive motivation for why the privacy debate is still alive for private citizens and institutions, and we investigate the privacy concept.
This paper proposes our vision of the privacy ecosystem, introducing privacy dimensions, the related users’ expectations, the privacy violations, and the changing factors. We provide a critical assessment of the Privacy by Design paradigm, strategies, tactics, patterns, and Privacy-Enhancing Technologies, highlighting the current open issues.
We believe that promising approaches to tackle the privacy challenges move in two directions: (i) identification of effective privacy metrics; and (ii) adoption of formal tools to design privacy-compliant applications.
\end{abstract}


\input{TPS-intro} 
\input{TPS-privacy}
\input{TPS-PbD}
\input{TPS-Gap}
\input{TPS-Conclusions}

\section*{Acknowledgments}
The author would like to thank Dr. Marco Benedetti and the Applied Research Team (IT Department, Bank of Italy) for support.

\bibliographystyle{IEEEtran}
\bibliography{TPS_Bib}

\end{document}

%% file: TPS-intro.tex
\section{Introduction}
\label{intro}

Let us start finding out how people feel about their privacy and the related technology issues. Do they really care about on-and off-line privacy? If we examine opinion polls and reports on the matter, we will not surprisingly find a generalized level of concern. According to a survey of the European Union Agency of Fundamental Rights \cite{FRA20}, $41$\% of Europeans are unwilling to share personal data with private companies; furthermore, this willingness depends on the data type. $24$\% of users are not able to set privacy settings on their apps, and $33$\% do not read terms and conditions when accessing online services. 
\begin{comment}
It emerged from an Ipsos survey that Americans are concerned with data privacy and security and ask for solid regulation to avoid unwanted and unauthorized data usage by technology companies \cite{Ipsos21}. 
\end{comment}
Strong demand for regulation over Big Tech emerges in the Amnesty International poll \cite{Amnesty}, where people admit the fear of the loss of control over their data. 
On the other hand, even if people claim to be concerned about their data privacy, they act as they do not care too much, sharing data for small rewards that often are just window dressing \cite{acquisti2013privacy}.

Undeniably, from time to time, we ask ourselves: are companies targeting me? And government and institutions? Is the personalization of my digital experience influencing me politically and socially? Are we dealing with our second self, one that has a ``digital biography [...] an unauthorized one, only partially true and very reductive'' \cite{solove2004digital}?		
As it appears from this brief analysis, new devices and the way we interact with them have changed the ``right to be left alone'' \cite{brandeis1890right}, just as the advent of photography and newspapers did in 1890. Thus, defining privacy merely as the desire for solitude and preserving reputation is not enough anymore: privacy embraces various aspects of our inner reality and how we interact within the community. Furthermore, privacy is highly dependent on time and context \cite{nissenbaum2020privacy}.
\begin{comment}
\smallskip \noindent \textbf{Motivation}. 
\end{comment}

Let us consider two use cases to understand how the perception of privacy varies accordingly to societal and technological changes. 
The first use case is 
%taken directly from newspapers and is 
related to the adopted measures for the containment of Covid-19. This unpredictable event has forced us to expand the borders of what is meant for privacy. The advent of contact tracing applications and the recent introduction of the Digital Covid Certificate to facilitate safe free movements inside the EU \cite{Pass2021} ask us to redefine our privacy concept for the sake of preventing the spread of the pandemic. Never as in this particular moment in history, data has no borders \cite{UNHR}.
Although contact tracing apps could represent a tool for the containment of Covid-19, they do not have the expected adoption. The lack of understandable and reachable information about the behavior of tracing apps does not encourage the adoption of them among the population. The mechanisms of terms and conditions need to be complemented with the ability to build trust between technology providers and users, offering a fair assessment of potential harms rather than blind assurances on the preservation of privacy \cite{national2007engaging}.
Moreover, legal privacy frameworks do not follow privacy threats as they emerge and are seldom flexible enough to follow the dynamic definition of privacy. Regulation is reactive, not proactive \cite{klitou2014privacy} and it insists on regulating the use of technology, not its development.

We move from private citizens to institutions for our second use case: privacy assumes a pivotal role in the execution of the institutional functions of a Central Bank \cite{iezzi2020practical}, \cite{bellomarini2021financialdata}. 
The Bank of Italy acts as the Italian supervisory authority, ensuring the proper management of intermediaries and compliance with the rules and regulations of those subject to supervision. Furthermore, it is responsible for the overall stability and efficiency of the financial system.
The Bank of Italy is also active in economic research and statistics to provide support for the decisions; it offers access to microdata via the Research Data Center to academia, National Statistics Officer, and other institutions, exclusively for research purposes. 
The Bank of Italy owns data from many business entities such as banks, intermediaries, and individuals for all these institutional goals. Among these valuable data assets, we find balance sheets of banks and intermediaries, payment systems data, datasets about Italian companies, datasets about loans granted and guarantees issued to households and firms, datasets about individuals subjected to revocation of payment cards because of missed payments, and many others. 
Furthermore, methodologies for processing these assets are changing, and new analysis opportunities arise, guided by technological advances, such as the availability of public cloud resources or the adoption of Machine Learning algorithms.
At the heart of a Central Bank’s sensitive activities, the trade-off between privacy, transparency of processes, the accuracy of analysis, the efficacy of IT systems, and compliance with regulatory frameworks such as GDPR is unavoidable. Data privacy should be enhanced by efficiently developing and evolving the underlying IT systems, such that regulatory measures to preserve privacy are not too demanding, invasive, and ``tech-unfriendly.''

Privacy by Design \cite{cavoukian2012operationalizing} tries to supply this necessity: the methodology was introduced to bridge privacy regulation and technology, giving dignity to privacy requirements and enhancing it at the same level as any other critical requirement. The privacy by design paradigm clarifies that the design of privacy-preserving IT services encompass the technical design and choice of effective Privacy-Enhancing Technologies (PETs) and the design of business and management processes. An integrated approach is needed to adhere to the complex and multidimensional definition of privacy. Ann Cavoukian \cite{cavoukian2012operationalizing} suggests that ``Privacy-Enhancing Technologies are necessary but insufficient to protect privacy and provide a foundation of trust well into the future.'' On the other hand, PETs could be not only an enabler but also a bearer of this trust. PETs are tangible and potentially understandable by design: they are associated with the applications that compulsively interact with the user; furthermore, they are easily accessible by large sections of the population, more than any other legal framework. Despite this, it is not straightforward how to match PETs to privacy requirements, measure the fulfillment of these requirements, and foster trust in PETs among citizens.  

When it comes to integrating Privacy by Design in enterprise contexts, various factors contribute to its limited adoption:
\begin{itemize}
    \item Regulation is usually on data providers and not on service providers, and this results in limited incentives in its adoption \cite{klitou2014privacy}.
    \item There is a lack of concrete guidelines and privacy metrics that could constitute an assessment tool for the considered use case and a point of connection with the existing legal framework.
    \item Technology and privacy regulation speak different languages that evolve with different speeds, and the definition of privacy evolves at the speed of technology adoption, societal changes, and unpredictable events. 
%\item Even if privacy regulation is reactive to new technological vulnerabilities, it still acts as a backward framing component for the definition of privacy.
\end{itemize}

Some unique opportunities for investigation arise in this background:
(i) the multifaceted definition of privacy and related concerns; (ii)
the technical means to guarantee the appropriate definition of privacy;
(iii) the gap between regulation and technology.

\smallskip \noindent \textbf{Contribution}. Motivated by these thoughts, we propose our vision of the privacy concept and critically review the Privacy by Design paradigm. Moreover, we provide a perspective on addressing the major current open problems. 

We provide the following contributions: 
\begin{itemize}
    \item To motivate our vision, in Section \ref{privacy}, we review the current definitions of privacy and propose a privacy ecosystem, identifying privacy dimensions, the related users’ expectations, the privacy violations, and the changing factors.
    \item In Section \ref {PbD}, we discuss the Privacy by Design paradigm and its interaction with privacy patterns and PETs, highlighting the current open issues.
    \item In Section \ref{gap}, we outline those that, in our opinion, are the primary directions for future research. 
\end{itemize}		 	 	 	

We found many valuable contributions to the systematization of privacy definitions and related practical challenges in literature. %To mention a couple, the authors in \cite{HippoDB} identify the problems in designing Hippocratic databases and provide approaches to overcome these issues. Bertino, in the recent paper \cite{Bertino}, proposes a discussion of the challenges related to privacy preservation in the context of 5G, IoT, Big Data, and Machine Learning.    	 	 	 		
In contrast to the other related works, this vision paper intends to shed light on the current panorama of the privacy concept and explore challenges in a broader and more general context.

%% file: TPS-privacy.tex
\section{What is Meant for Privacy}
\label{privacy}

The etymological meaning of privacy derives from the Latin \emph{privatus}, which originates from \emph{privus}: it means individual, single, personal but also peculiar, unique. Indeed, privacy refers not only to one's body or material possessions but also to the uniqueness of thoughts, behaviors, and emotions \cite{Clarke1}.
Privacy is a complex concept, and giving a comprehensive definition of privacy is an arduous task.
\begin{comment}
its nature initially arises in social, psychological, philosophical, juridical, and political sciences, and its technological dimension has become prominent only in the last years. 
\end{comment}

The "right to be left alone," as in \cite{brandeis1890right}, refers to the concept of privacy \textbf{as restricted access} to personal space, or with the words of Roger Clarke \cite{Clarke2}:  "privacy is the interest that individuals have in sustaining 'personal space,' free from interference by other people and organizations." We can revise a dual connotation for this personal space whenever this space is merely physical or not.

If we intend the personal space as material, privacy pertains to the faculty and the capacity to prevent intrusion. Privacy is the ability to avoid "the observation of [...] one's home, and personal belongings" \cite{schoeman1992privacy}, or, as in Clarke's vision \cite{Clarke1}, as the "bodily privacy," the guarantee of "the integrity of the individual's body."

Referring to the immaterial aspects of an individual, privacy is often depicted \textbf{as autonomy within society and human freedom} \cite{national2007engaging},\cite{schoeman1992privacy}, and it constitutes "an indispensable feature of a democracy where an individual maintains his identity while contributing to their civic duty" \cite{cohen2012privacy}.
This definition of privacy gives the floor to the view of privacy \textbf{as a human and constitutional right} \cite{Clarke1}, \cite{terstegge2007privacy}, and a debate arises around this definition of privacy \cite{Forbes}. 

The ability to protect our personal and emotional spheres and the information contained therein implies an idea of privacy \textbf{as control} \cite{belanger2011privacy} "over how information flows"\cite{Boyd}. As argued by Westin \cite{westin1968privacy}, privacy is "the claim of individuals, groups, or institutions to determine for themselves when, how, and to what extent information about them is communicated to others." 
In the work of the social psychologist Irwin Altman, privacy is defined as "selective control of access to the self or to one's group" \cite{altman1977privacy}: this leads to two important considerations: (i) privacy is not only a matter of disclosure but of selective disclosure of information; (ii) the personal space is not only one's body or one's belongings but also one's group, where the individual takes one of his multiple identities and discloses information based on the role he assumes in that group. 

In the digital economy, information is power, and the commodification of personal information and habits suggests that privacy could also be viewed \textbf{as an economic value} \cite{acquisti2009nudging}, \cite{acquisti2016economics}, \cite{acquisti2015privacy} or \textbf{as an exchanging good} \cite{posner1977right}. The authors in \cite{acquisti2013privacy} found that individuals fluctuate between two opposite directions, (i) the willingness to accept payments or other goods or products in exchange for their private information, and (ii) the willingness to pay to protect their private information. Not surprisingly, people are more willing to disclose their information for a reward rather than pay to protect their data: privacy is perceived as something that could be sold but only seldom bought. This is true even if people affirm to be concerned about their privacy: this is known as the \emph{privacy paradox}, which was first introduced by Barry Brown in 2001 \cite{brown2001studying}. In Brown’s study, privacy concerns contradict the willingness of people to share their data via store loyalty cards for little reward. 

Taking inspiration from \cite{national2007engaging}, we identify three main dimensions for privacy: 

\begin{itemize}
    \item \emph{Physical dimension}, where privacy is the capacity to restrict access to the individual and his physical space selectively.
    \item \emph{Informational dimension}, where privacy is control over personal information and the ability to manage personal information as an economic value.
    \item \emph{Decision dimension}, where privacy is the right and the freedom to act autonomously within society. 
\end{itemize}

\begin{figure*}[ht!]
  \centering
    \includegraphics[width=0.75\textwidth]{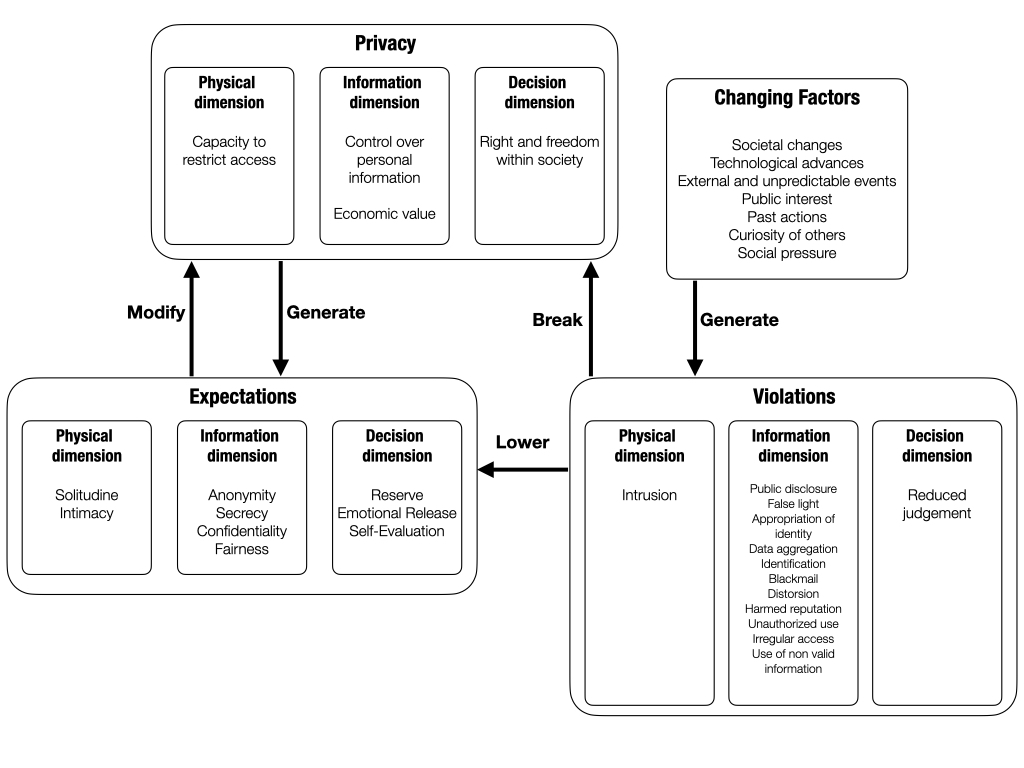}
  \caption{The privacy ecosystem.}
  \label{fig:ecosystem}
\end{figure*}

\subsection{Privacy Expectations}
\label{expectations}

In \cite{national2007engaging}, the authors claim that ``as a society, we share some basic expectations of privacy.'' We add that we have expectations also as individuals. A privacy expectation is the expected result of the fulfillment of the definition of privacy in a given dimension, i.e., what an individual expects if he can exert his privacy in each dimension given the context. 
%We consider privacy as a multilevel construct, as in \cite{rao2020types}, with four levels of fulfillment: Desired, Predicted, Deserved, and Minimum.

For the physical dimension, the expectations for privacy are \cite{westin1968privacy}: (i) \emph{solitude}, that is, physical inaccessibility; (ii) \emph{intimacy}, which is restricted accessibility of the self to a small group. 
For the informational dimension, privacy should enable \cite{westin1968privacy}, \cite{gavison1980privacy}: (i) \emph{anonymity}, that is the ability of not being identified; (ii) \emph{secrecy}, that is closure in terms of knowledge; (iii) \emph{confidentiality}, that is selective disclosure of information based on rules and context; (iv) \emph{fairness}, that is the expectation that disclosed information would be used solely for the intended purpose. 
For the decision dimension, the expectations are \cite{westin1968privacy}: (i) \emph{reserve}, that is a ``caution'' in sharing information in order to avoid unpleasing invasion;  (ii) \emph{emotional release}, that is the relief from social rules and roles; (iii) \emph{self-evaluation}, that is the internal talk about oneself and his environment. We remark that reserve is different from confidentiality: reserve is a psychological concept, and it is intended like not regulated: it is only ``protected by the willing discretion of those surrounding him'' \cite{westin1968privacy}. On the other hand, confidentiality can be achieved by legal means too.  

Interestingly, expectations for privacy are sometimes confused with the definition of privacy itself \cite{smith2011information} and, in the work of Westin \cite{westin1968privacy}, and Gavison \cite{gavison1980privacy}, they are often referred to as states or functions of privacy. We prefer to consider them as expectations due to the ephemeral and subjective nature of their definitions. Furthermore, some of the expectations belonging to one privacy dimension could easily be assigned to another dimension.
%e.g., intimacy could also be an expectation of informational privacy. The individual discloses information based on the role he assumes in a social group. 

\subsection{Privacy Violations} 

All the presented definitions of privacy require that individuals are conscious of the consequences of its violations, but as far as these violations do not occur, most users do not think about privacy as a concern \cite{spiekermann2008engineering}.
For privacy violations, we intend the actual effects of the total or partial unfulfillment of the definition of privacy for a given dimension that breaks almost one of the expectations for that dimension. 
There have been many attempts at classifying privacy violations; this paper considers only some of the available taxonomies, referring each violation to the identified privacy dimension. We see that the distinction between the informational and the physical dimension of privacy, i.e., privacy as control and privacy as a personal space, respectively, was already present in \cite{posner1977right}, where Posner individuates four types of violations. Three violations involve the informational sphere of the individual: (i) public disclosure of private facts; (ii) false light and harmed reputation; (iii) appropriation of identity. The last one, intrusion, involves the physical dimension of privacy, breaking the expectation of solitude. 
Smith et al. in \cite{smith1996information} extend Posner's harmful activities to the informational dimension, including internal and external unauthorized use, use of inaccurate or not valid information, and irregular access to personal information. Regarding the decision dimension of privacy, the authors also included reduced judgment in decision-making due to automated procedures or incorrect or partial information.
Solove's taxonomy in 2006 \cite{solove2005taxonomy} poses great attention to violations to the informational dimension of privacy, grouping them in the three main phases of information-related activities: collection, processing, and dissemination. Besides the already mentioned harmful activities, we find data aggregation, identification, breaking the expectation of anonymity, insecurity, blackmail, and distorsion.

\subsection{Changing Factors}

Privacy cannot be defined in an unambiguous and general way, as argued in \cite{morton2012privacy}: it depends on the perception of what is considered private by the individual and by the society, that changes over time. 
The mobile nature of privacy has been introduced by Altman \cite{altman1977privacy} and later by Palen \cite{palen2003unpacking}: privacy is seen as a dynamic and dialectic boundary regulation process. Privacy is not rule-based, and it is a search for balance between openness and closure depending on boundaries that are influenced by experiences, context, technologies.
Analogously, also privacy expectations are time-varying. In 1967, Justice Harlan introduced this aspect in Katz v. United States \cite{Legal} and his reasonable expectation of privacy test. It distinguishes between the subjective expectation of privacy, what one feels as private, and the objective expectation of privacy, what society recognizes as private and could be protected by law. 
Even if we do not focus on the legal aspects of privacy, we agree that privacy expectations are a mixture of subjective and objective factors that evolve with social norms and technological changes.

We identify seven main drivers \cite{national2007engaging}, \cite{westin1968privacy}, \cite{palen2003unpacking}
: (i) societal changes: what we consider private today could not be private anymore tomorrow; (ii) new technological advances that enable new vulnerabilities and modify our perceived boundaries; (iii) external and unpredictable events, e.g., a pandemic that may require our personal data to stop the spreading of the virus; (iv) public interest, e.g., national security or medical research; (v) past actions; (vi) curiosity of others, and (vii) social pressures.

\subsection{Putting It All Together}

The result is that privacy appears as a foggy abstraction that embraces individuals, their private information, rights, freedom, abilities, and comprehension of all involved concepts. Though many efforts have been devoted to privacy regulation, the puzzling definition of privacy that results from our analysis confirms that "the value of privacy, while not entirely arbitrary, is highly malleable and sensitive to non-normative factors" \cite{acquisti2013privacy}.
In the remainder of the paper, we will focus on the informational dimension of privacy. The current scenario is the one introduced in \cite{solove2005taxonomy} and enriched in \cite{GDPR}, identifying three entities as follows: (i) data subject, that provides personal data, (ii) data controller, that establishes the means and the purpose of data collection, and (iii) data processor, that processes personal data. 
In this scenario, collecting and processing data has become increasingly complex due to the principle of data maximization, advancement in new analysis models, and the availability of high computational resources. 

Transparency about data usage and computation seldom lacks or is not sufficient. Moreover, it is challenging to calibrate communication about privacy implications of data processing to all groups of individuals, as they are characterized by different ages, gender, culture: privacy is also an educational and cultural question.

In Fig. \ref{fig:ecosystem}, we represent the privacy ecosystem and the interrelations among privacy dimensions, expectations, violations, and changing factors. We see that changing factors generate new privacy violations that break the notion of privacy and lower individuals’ expectations. The notion of privacy is strongly interconnected with expectations, in a two-way influence. 

%% file: TPS-PbD.tex
\section{Matching Technology to Regulation: the Privacy by Design Paradigm}
\label{PbD}

Regulatory frameworks alone are not sufficient to follow the evolution of the concept of privacy: as stated by Lessig \cite{lessig2009code}, like it or not, ``code is law,'' and software developers ``are increasingly lawmakers.''
In 2009, Anna Cavoukian \cite{cavoukian2012operationalizing} introduced Privacy by Design to formalize the mapping between the legal corpus of privacy, IT systems and infrastructures, and the organizational and management context. The approach is holistic because it covers not only the technical aspects but also enterprise processes. 
The Privacy by Design framework states that the privacy goals and related requirements should be proactively considered in the early stages of the design and development of IT systems, and it provides a set of principles and guidelines that should act as a driver for the design of privacy-aware applications. 

Privacy goals and requirements for the developers are conceptually related to the general privacy expectations of the users towards the IT system for the informational dimension, as we described in Section \ref{privacy}. However, they are far more concrete: they are system quality attributes deriving from assessing potential violations. For example, among privacy goals, we find unlinkability, unobservability, pseudonymity \cite{colesky2016critical}, \cite{fischer2001security}. 

This comprehensive approach to privacy integration should be maintained across all the application lifecycle. It implies that developers and users work together to define what is private and not and the best approach to fulfill the privacy requirements. 
The Privacy by Design framework covers three main aspects: 

\begin{itemize}
    \item \emph{Principles} \cite{cavoukian2012operationalizing}, that describe practices for privacy-compliant IT systems;
    \item \emph{Strategies} \cite{hoepman2014privacy}, that implement the principles and provide the high-level description of structural realization of privacy requirements;
    \item \emph{Tactics and related patterns} \cite{colesky2016critical}, that help to achieve privacy strategies and are more technical-oriented.
\end{itemize}

\subsection{Principles, Strategies, and Tactics}

Privacy by Design \emph{principles} mainly deal with the development of IT systems; the seven principles range from ``stating the necessity of privacy as default setting'' to ``preserving usability and functionality of IT system while preserving privacy.''
For the sake of completeness, we remind that these privacy principles do not deal specifically with the data collection, processing, and dissemination processes; some interesting principles for the data issues are given within, e.g., the GDPR \cite{GDPR}.

For the design phase of an IT system, Hoepman in \cite{hoepman2014privacy} introduces \emph{strategies} as a general instrument to engineer Privacy by Design principles. 
Strategies derive from privacy violations and conceptualize how to avoid them: they describe a general approach to achieving a fixed privacy goal. Hoepman starts from a specific use case and then derives eight general strategies in the realm of Privacy by Design engineering. Four strategies are said to be data-oriented because they involve the data processing phase; on the other hand, four strategies are said to be process-oriented as they deal with the organizational processes. 

\emph{Tactics} constitute a specialization of the strategies \cite{colesky2016critical} that provide hints for the technical solution implementation. There is more than one tactic associated with each strategy. 
Strategies and tactics are at a level of abstraction that comes before privacy patterns and can address privacy requirements since the early design phase.
For instance, the HIDE strategy is defined as a general guideline: ``Any personal information that is processed should be hidden from plain view \cite{hoepman2014privacy}.''
One tactic for the HIDE strategy is OBFUSCATE, which aims at ``preventing understandability of personal data to those without the ability to decipher it,'' a suggestion for the adoption of Encryption. In Table \ref{tab:PbD}, details about strategies and tactics are given.

\begin{comment}
\begin{itemize}
    \item Proactive not reactive, preventative not remedial: it anticipates privacy harmful events rather than resolving them after they occur;
    \item Privacy as default setting: data privacy should be assured automatically;
    \item Privacy embedded into design: it encourages the practice of considering privacy as a fundamental requirement and not as a burden for the usability of the system;
    \item Positive-sum, not zero-sum: privacy requirements fulfillment should not come at the cost of other functional and nonfunctional requirements;
    \item End-to-end security, full lifecycle protection: the IT system should be protected throughout all its lifecycle;
    \item Visibility and transparency, keep it open: every component of the IT system should be transparent to the users and providers;
    \item Respect for user privacy, keep it user-centric: developers should design and implement the IT system by providing features that satisfy the users’ privacy expectations. 
\end{itemize}
\end{comment}

\begin{comment}
\subsection{Strategies and Tactics}
\label{strategies}
\end{comment}

\begin{table*}[ht!]
\caption{Strategies, Tactics, and Patterns \cite{colesky2016critical}, \cite{hoepman2014privacy}, \cite{Pattern1}, \cite{Pattern2}}
\label{tab:PbD}
\centering
\resizebox{\linewidth}{!}{%
\begin{tabularx}{\textwidth}{|l|X|X|X|X|}
\hline
Type & Strategy & Description & Tactics & Patterns \\ 
\hline
\multirow{4}{*}{Data-oriented} & Minimize & The amount of personal information that is processed should be minimal. & Exclude, Select, Strip, Destroy & Attribute-Based Credentials, User data confinement pattern, Aggregation Gateway, Dynamic Location Granularity \\ 
\cline{2-5}
 & Hide & Any personal information that is processed should be hidden from plain view. & Restrict, Mix, Obfuscate, Dissociate & Anonymity Set, Protection Against Tracking, Strip Metadata, Pseudonymous Identity, Onion Routing, Anonymous Reputation-based Blacklisting \\ 
\cline{2-5}
 & Separate & The processing of personal information should be done in a distributed fashion whenever possible. & Distribute, Isolate & Aggregation Gateway, Anonymous Reputation-based Blacklisting \\ 
\cline{2-5}
 & Abstract (former Aggregate) & Personal information should be processed with the least possible detail in which it is (still) useful. & Summarize, Group & Anonymity Set, Aggregation Gateway, Trustworthy Privacy Plug-in, Dynamic Location Granularity \\ 
\hline
\multirow{4}{*}{Process-oriented} & Inform & Data subjects should be adequately informed whenever personal information is processed. & Supply, Notify, Explain & Privacy-Aware Network Client, Privacy Color Coding, Privacy icons, Privacy Dashboard, Layered Policy Design, Policy Matching Display \\ 
\cline{2-5}
 & Control & Data subjects should have agency over the processing of their personal information. & Consent, Choose, Update & Privacy Color Coding, Data Breach Notification Pattern, Privacy Dashboard, Policy Matching Display, Private link \\ 
\cline{2-5}
 & Enforce & A privacy policy compatible with legal requirements should be in place and should be enforced. & Retract, Create, Maintain & Federated Privacy Impact Assessment, Sticky Policies, Obligation Management \\ 
\cline{2-5}
 & Demonstrate & Be able to demonstrate compliance with the privacy policy and any applicable legal requirements. & Uphold, Audit, Log, Report & N/A \\
\hline
\end{tabularx}
}
\end{table*}

\subsection{Privacy Patterns and Privacy Enhancing Technologies (PETs)}
\label{PETs}

When it comes to the development of Privacy by Design strategies and tactics, we need to resort to privacy patterns and Privacy-Enhancing Technologies (PETs).
Privacy patterns \cite{diamantopoulou2017supporting} represent specific solutions to an implementation problem. Indeed, privacy patterns are recurring schemes that conceptualize the components and the relations between them to achieve the overarching tactic and strategy. The idea is to provide the developers with a common framework to standardize privacy solutions and implement legal requirements. In \cite{Pattern2}, there is a repository of common patterns categorized by tactics; a similar collection can be found in \cite{Pattern1}, where patterns are tagged by the related strategy. A summarization of some of the main patterns is in Table \ref{tab:PbD}.

PETs can be defined as the set of ``ICT measures protecting informational privacy by eliminating or minimizing personal data thereby preventing unnecessary or unwanted processing of personal data, without the loss of the functionality of the information system'' \cite{danezis2015privacy}. PETs are the technical implementation of a given pattern. 
For example, in the case of the couple HIDE-OBFUSCATE during data processing, the pattern ``Anonymity Set'' can be implemented by Onion Routing. 

The desiderata are to move trust from service providers to the chosen technology, given that there exists a specific understandable and reliable mapping between a regulatory standard and a  particular PET. Trust can be perceived as an indicator of the confidence from a trustor, typically the data subject, to trustees identified by the data controller and data processor, that should act as ``privacy keepers.'' Trustees are supposed to possess skills to protect the data subject's privacy, care about his/her privacy, and be loyal to written and unwritten rules. 

In this sense, PETs should act as an enabler of the concept of privacy as control by default for the data subjects, providing them an instrument to exert the ``informational self-determination'' \cite{fischer2001security} and to enhance trust in the data controllers, rather than relying only on regulation. 
Furthermore, as clearly stated by \cite{Royal}, the role of governments and regulators is pivotal to the diffusion of best practices for the preservation of privacy and research in the area of PETs: their adoption encourages sharing of data with the academic community, enhances competitiveness, and strengthens the trust of individuals in technology.   

\subsection{Challenges of Privacy by Design}

Although the introduction of Privacy by Design and PETs has represented the first step towards a common ground between regulation and technology, these instruments appear to fail to bring efficacy and adapt to the evolving definition of privacy and users’ concerns. The user is not always aware of what Privacy by Design means and cannot verify if an application is compliant with what Privacy by Design promises to deliver. It results that this paradigm strives at becoming an instrument of trust for users.   
Privacy by Design is still a vague concept, and even if high-level guidelines are given, they are not concrete and subjected to interpretation. In the remainder of the section, we identify two main challenges in real-world Privacy by Design adoption. 

\textbf{Operationalizing Privacy by Design}. As highlighted in \cite{cavoukian2012operationalizing}, there is a recurrent mismatch between Privacy by Design guidelines and software engineering practices. Generally, privacy requirements are fixed from the beginning, and the development follows the common waterfall methodology. Different approaches are proposed in \cite{PRIPARE} and \cite{hoepman2018privacy}, where the authors briefly refer to an iterative methodology. The use of iterative agile developers paradigms entails moving functional requirements: this has a dual result (i) the need to change requirements during the iterations may affect the privacy goals that were sustained by the initial solution; (ii) on the other hand, iterative development practices well commit to the intrinsically dynamic concept of privacy. 

\textbf{Do not abandon (your) PET}. The diffusion of PETs encounters some difficulties since their usage seems complex and expensive; moreover, the common belief that applying PETs makes the application more privacy-centric and less user-centric does not help. 
This could be true to a certain extent if PETs are seen as plug-in components that are added at the end of the design process of the IT system or as components to be integrated into legacy systems: thus, PETs should be part of the Privacy by Design process, reducing the structural costs of implementation with respect to a post-integration approach. The lack of standardization of PETs and the inherent interoperability problems are other obstacles that sum up to the already mentioned. 
Furthermore, there is an apparent lack of an economic incentive to the introduction of PETs in IT systems: as argued in Section \ref{privacy}, the privacy paradox affects the users and does not always make privacy a vital requirement for them. On the other hand, the implementation of PETs is under the service provider's responsibility, but it is not a priority unless there is a legal requirement.

%% file: TPS-Gap.tex
\section{Mind the Gap(s)}
\label{gap}

There is a substantial gap between privacy regulation and the introduction of privacy principles, guidelines, and technology in the development of privacy-aware IT systems. In this section, we highlight two main weaknesses in bridging privacy regulation and technology: (i) the lack of privacy metrics that would allow a more rigorous matching between technological privacy tools and legal requirements; (i) the lack of a formal privacy framework as a translator of legal requirements into technological solutions. For each of these issues, we suggest some directions from available literature to address the challenges. Note that our analysis is far from complete: the aim here is to point out the ones that could be the most promising in our vision.  

\subsection{Measuring Privacy and the Information-Theoretic Approach}
\label{measures}

The authors in \cite{nissim2017bridging} argue that a critical step in the adoption of Privacy by Design and PETs is the demonstration that these instruments can satisfy relevant legal requirements; this is critical because there is the need to reconcile the legal and the mathematical approaches that are commonly used for privacy measurement. 
Indeed, Privacy by Design suggests that a priori strategy should be applied \cite{danezis2015privacy}: we choose the objective function or the metric to minimize, that is related to privacy goals and expectations, we adjust the parameters of the obtained model to fulfill the fixed privacy goals, and then we choose the PETs to apply. Thus, a well-posed metric that assesses the privacy level is central to deciding the most suitable PET.  
The metric to minimize should also account for the utility of the solution, which is a conflicting goal with privacy. In other words, the metric should assess the so-called \emph{privacy-utility trade-off}. 
Moreover, privacy preservation is not guaranteed in the presence of system composition: if a system satisfies a specific property of privacy, this property may not be fully satisfied if the system is connected to another system. This fragility of composition has been proven for differential privacy in the work of Kairouz et al. \cite{kairouz2015composition}, where they prove the composition theorem: sequential querying of differentially private mechanisms degrades the overall privacy level. It is desirable that the privacy metric accounts in a reliable way for the composition of privacy countermeasures.
In \cite{wagner2018technical}, a complete review of technical privacy measures are provided, under four dimensions: (i) adversary model; (ii) data sources, that the IT system and related PETs protect; (iii) inputs needed for the metrics; (iv) output, that characterizes the privacy attribute or expectation that is captured by the metric. Concerning the latter dimension, the authors remark on the multifaceted nature of privacy, arguing that just one measure is not sufficient to describe the whole concept of privacy. 

\smallskip \noindent \textbf{Statistical Disclosure Control and Differential(ly measuring) privacy}. In the context of Statistical Disclosure Control, various tractable, from a mathematical point of view, metrics have been proposed; these are metrics that return the properties of disclosed data and the similarity between them; the most popular are: (i) $k$-anonymity; (ii) $l$-diversity; (iii) $t$-closeness \cite{wagner2018technical}.
\begin{comment}
\begin{itemize}
    \item $k$-anonymity \cite{sweeney2002k}, where each tuple of key attributes, i.e., quasi-identifiers, is indistinguishable related to no fewer of other $k$ tuples, or, in other words, we can group tuples in equivalence classes with at least $k$ indistinguishable rows with key attributes;
    \item $l$-diversity, where each key attribute should have at least $l$  ``well-represented'' \cite{machanavajjhala2007diversity};
    \item $t$-closeness, where the distance between the distribution of key attributes in any equivalence class does not exceed a threshold $t$ \cite{li2007t}.
\end{itemize}
\end{comment}
A popular technique that deals with the concept of indistinguishability is Differential Privacy (DP), introduced in \cite{dwork2006differential}. DP portraits what it means to be private in a data publishing and sharing context. The associated metric is the distance between adjacent databases that differ on one individual. 
Promising research areas address the opportunity of extending the classical DP metric to broader domains, not necessarily that of the statistical databases, as in the work of \cite{chatzikokolakis2013broadening}.
Another promising direction is provided in \cite{samaratifinal}, where the differential privacy parameters are chosen according to the legal requirements: this leads to considering DP parameters as metrics for identifiability. 
The authors derive two bounds that developers could use in practice to choose DP parameters for Machine Learning scenarios that are resistant to membership inference attack and to DP adversary, i.e., maximum posterior belief and expected membership advantage.

\smallskip \noindent \textbf{An information-theoretic approach to measuring privacy}. This approach derives from the assumption that communication and cryptography are two sides of the same coin: if in communication we aim at extracting information from data, in privacy-preserving contexts, we aim at disclosing a quantity of data that does not allow to estimate undisclosed data \cite{calmon2015information}. Under this umbrella, we mainly find entropy-based metrics and privacy-utility bounds useful to design IT systems under a fixed privacy budget. 
Indeed, Shannon’s entropy is a measure for the level of anonymity observable by an attacker, or, in other words, is the number of additional bits that would allow an adversary to identify an individual in a set \cite{serjantov2002towards}. Since this first metric, many attempts have been made to characterize privacy following the information-theoretic approach. 

Vincent Poor et al. in \cite{sankar2010information} propose using the rate-distortion theory to derive a privacy-utility trade-off built on a database model with an arbitrary number of public and private variables. 
In \cite{alfalayleh2014quantifying}, the information-theoretic approach is used for measuring the disclosure risk, i.e., the risk of identifying an individual in a database or the risk of disclosing a confidential value, that could be considered as a measure of privacy as well, enabling the evaluation and the comparison of the most common statistical disclosure control methods, such as sampling, noise addition, and query restriction.
%In \cite{calmon2015fundamental}, a condition for disclosing a non-trivial amount of information while not disclosing private information is given via the Principal Inertia Component method. At the same time, this allows determining the trade-off between disclosure of useful information and protection of private information.
The use of entropy rate for evaluating mobility profile in location-based services is presented in \cite{rodriguez2015entropy}, fulfilling the necessity to take into account processes with memory: the correlation between successive locations is not negligible in profiling. 
A unifying framework for all the various entropy-based privacy metrics is provided in \cite{rebollo2013measurement}, based on the assumption that the estimation error of an adversary while trying to disclose information from a privacy-preserving IT system could act as a valid privacy metric. 

The information-theoretic approach has also been applied to Differential Privacy; an example can be found in \cite{mir2012information}, where the author %, similarly in \cite{sankar2010information},
uses the rate-distortion problem to derive privacy-utility bounds for differential privacy. Furthermore, a strong connection between differential privacy mechanisms and the maximum entropy principle is demonstrated. 

\subsection{The Promise of Formal Approaches: Languages, Logics, Contextual Integrity \& Friends}
\label{languages}

Besides Privacy by Design best practices, developers are also asked to face the overall design of the architecture of IT systems, the identification of the actors involved and of the information flows, and only then, they choose the suitable PETs. In our vision, formal frameworks ensure that the technical implementation corresponds to the defined privacy model and satisfies the legal requirements and the users’ expectations. A rigorous but flexible methodology to drive the developers in this choice, given privacy requirements, is lacking and is not covered by Privacy by Design literature.  
Moreover, there is an increasing need for transparency, verification, and explainability of privacy-aware IT systems. Disclosing the source code or recurring to the use of open source is indeed a best practice, but it is not sufficient because it does not provide a proper verification instrument for the legal experts, since the gap between legal and technical languages is too big \cite{diver2017opening}.  We remark that checking a source code's correctness and compliance to a given legislation corpus constitutes a notable effort even for an expert developer. 		 
Formal methods increase trust in the IT systems because potentially both users and developers can verify the overall system's compliance. 

%In \cite{nissim2017bridging}, the authors recall the known approach in the computer science community of formalizing and assessing privacy goals or expectations as a ``game, or a thought experiment, in which an adversary attempts to exploit a computation to learn protected information.'' The game should be carefully designed in all its components, i.e., the identification of the adversary and the boundaries in which the adversary can operate, when the adversary wins the game, or otherwise when the system is assessed as privacy-preserving. However, it is not straightforward to understand how this formalization can be applied to tie regulation to IT systems in a structured way. 

%\begin{comment}
Inspired by the taxonomy in \cite{tschantz2009formal}, we propose two main families: 
\begin{itemize}
\item \emph{Formal models}, to represent the behavior of the system and its environment, and related interactions.
\item \emph{Formal logics and languages}, to describe dimensions of privacy, expectation, policies, and to reason about the model.
\end{itemize}

\smallskip \noindent  \textbf{Formal models}. A formal privacy model is described in the work of Simone Fischer-H\"ubner of 1998 \cite{fischer1998formal} and then refined in \cite{fischer2001security} in 2001. The author presents a task-based privacy model, defined as a state machine model, that enforces a fixed privacy policy that implements a specific privacy regulation. 
\begin{comment}
The privacy model is inspired by the classical security model and involves the definition of (i) state variables, which are active subjects like processes, passive objects like files, attributes, tasks, and every other entity that describe the system state; (ii) invariants and constraints, that are the conditions under which the system respects the policy; (iv) state transition functions, which are the rules of change for the state variables. The introduced privacy model has been proven to be consistent with its axioms. 
\end{comment}
In \cite{diver2017opening}, the authors propose Petri Nets as an instrument to bridge legal requirements and norms to software processes in a way that is intelligible by both communities, and that frees the developer to recall and interpret legal norms. %Petri Nets is a standardized tool that, via two types of elements, states, and transitions, can represent concurrency, choices, and iterations. 
Petri Nets are sustained by a well-defined mathematical theory that allows a phase of static verification, a proof of the real-world functioning of the modeled IT systems. Petri Nets could be applied in the first phases of the Privacy by Design paradigm thanks to their high level of abstraction.
%\end{comment}

\smallskip \noindent \textbf{Formal languages and logics}. Formal languages to translate policies and privacy requirements into machine-readable forms have been proposed in the years. In \cite{anton2004financial}, a general method for analyzing privacy policies and decomposing them in actors, action works, and so on is proposed, while the recent work of Morel and Pardo gives a complete assessment of languages for machine-readable privacy policies. On the other hand, Guarda and Zannone in \cite{guarda2009towards} depict an overview of the available methods and constructs to aid privacy requirements engineering. 
A limitation of these formal languages for machine-readable policies is the lack of formal semantics that does not allow verification of compliance of the IT system with the implemented policies. However, this limitation is overcome in various proposals. In \cite{abe2016formal}, Z language is employed for modeling the system in its components and constraints and its behavior, while ProZ enables the validation. In \cite{stouppa2006formal}, a first-order relational language and an ontology enable to model the considered system in the context of the following privacy problem: ``none of the private data can be inferred from the information which is made public.'' The public information is constituted by a data view instance - queries and their answers - and additional knowledge to understand the data view. The privacy problem is investigated for relational databases and description logics, which are two decidable problems. In \cite{becker2010practical}, a declarative language with formal semantics to express privacy policies is introduced, %overcoming the limitation in the expressiveness of other languages and 
avoiding the need to specify the exact meaning of service behaviors, but leaving the room to specify it, 
``plugging'' any required logics to the purpose.
Jafari et al. \cite{jafari2011towards} propose a modal logic and a model checking algorithm for purpose-based privacy policies; the purpose can be inferred by the agents' actions in the model and their relationships, which can be modeled as an action graph. 

%Finally, suggestions about how to integrate mental models from human factors engineering with formal methods are given in \cite{houser2017formal}: mental models are helpful to express the interpretation of an individual of a target system, with special attention to privacy aspects, while formal methods on mental models can evaluate user expectations and the actual system properties. This encourages the spreading of privacy and security solutions for targeted slices of the population with peculiar needs. 

In the realm of formal languages and logic, particular attention is devoted to the \textbf{Contextual Integrity} (CI) framework. This framework poses its foundation on the concept developed by Helen Nissenbaum \cite{nissenbaum2020privacy}, where particular emphasis is given to privacy as a ``right to appropriate flow of personal information.'' It extends the concept of privacy as control, seen in Section \ref{privacy}, framing it as a concept regulated by context and related norms, purposes, and values: the promise of this approach is to take into account all the changes in users’ privacy expectations and acting as an instrument to convey the legal requirements. 
The CI framework introduces three concepts \cite{benthall2017contextual}: 
\begin{itemize}
	\item \emph{Context} or social spheres that describe the environment in terms of capacities,  roles, practices, norms.
	\item \emph{Contextual information norms} that make explicit users’ expectations and information flows, identifying five parameters: sender, recipient, and information subject, information types (topics, attributes), and transmission principles.
	\item \emph{Contextual ends, purposes, and values}, which are the attributes that characterize the meaning of a context.
\end{itemize}
CI is violated when the informational norms are not followed: this takes into account the societal and technological changes in the considered context, highlighting the need to update or introduce new informational norms, tuning the parameters. For example, updating a norm may involve the addition of a new recipient or of a new transmission principle. 
A formalization of CI is provided in \cite{barth2006privacy} where a logical framework is provided for reasoning over informational norms, encoding not only privacy expectations but also privacy legislation. The model encompasses evolving agents that communicate and interact with each other, giving rise to received messages, inferred attributes, and, more generally, ``execution histories, or traces.'' The temporal dimension of this execution history is integrated thanks to Linear Temporal Logic (LTL). Moreover, conjunction, disjunction, and implication are employed to allow policy combination and refinement. 
The CI framework is used in \cite{barth2007privacy} to model a business process and analyze it with respect to privacy compliance and business utility. CI is instrumental in defining business goals via purpose and privacy requirements via transmission goals. Privacy goals are verified by the use of Linear Temporal Logic (LTL), while the utility is verified by ATL$^*$.

%% file: TPS-Conclusions.tex
\section{Conclusions}
\label{conclusions}

Privacy is an old but ever-gold concept: it involves many facets that comprehend psychological, legal, sociological, and technological aspects. Citizens and institutions are both involved in the privacy debate. This vision paper contributes to the formalization of the privacy concept, placing it as a part of an integrated ecosystem. 
We have introduced and studied the interplay between Privacy by Design principles, strategies, tactics, patterns, and Privacy-Enhancing Technologies. Privacy by Design has become a valuable instrument for the transparent communication of the design phase of the IT system by the developers towards the users. 
However, it has open challenges; we discuss two of them, (i) the mismatch between Privacy by Design guidelines and software engineering practices; and (ii) the difficulty of adoption of PETs. 

Finally, we highlight that two strategies can be pursued to tie privacy regulation to privacy principles, guidelines, and technology in developing privacy-aware IT systems. One is the development of privacy metrics to allow a more rigorous matching between technological privacy tools and legal requirements. The other one is using formal privacy methods that may enable the translation of legal requirements into technological solutions and the verification of compliance with privacy policies.

We remark the importance of building a common privacy vocabulary between users, lawmakers,  and IT developers in the evolving privacy ecosystem: changing factors create new privacy violations and transform users’ expectations and perceptions of privacy. As argued in \cite{tschantz2009formal}, ``It is our responsibility as scientists and engineers to understand what can or cannot be done from a technical point of view on privacy: what is provably possible or impossible and what is practically possible or impossible. Otherwise, society may end up in a situation where privacy regulations put into place are technically infeasible to meet.'' Moreover, we add, vice versa: society may end up in a situation where emerging technical solutions could not be effectively regulated in time.